\documentclass[letterpaper,twocolumn,10pt]{article}
\usepackage{usenix-2020-09}

\usepackage{tikz}
\usepackage{amsmath}
\usepackage{xspace}

\usepackage{graphicx}
\usepackage{subcaption}

\usepackage{algorithm}
\usepackage{algorithmicx}
\usepackage{algpseudocode}
\algrenewcommand\algorithmicindent{1em}
\usepackage{booktabs}
\usepackage[load-configurations={abbreviations,binary},binary-units]{siunitx}
\usepackage{comment}

\usepackage{listings}
\definecolor{codegray}{rgb}{0.5,0.5,0.5}
\definecolor{codegreen}{rgb}{0,0.6,0}
\definecolor{backcolor}{rgb}{0.99,0.99,0.99}
\lstdefinestyle{pythoncode}{
    language=Python,
    backgroundcolor=\color{backcolor},
    basicstyle=\footnotesize\ttfamily,
    keywordstyle=\color{codegreen}\bfseries,
    commentstyle=\color{codegray}\itshape,
    stringstyle=\color{red},
    numbers=left,
    numberstyle=\tiny\color{gray},
    numbersep=8pt,
    breaklines=true,
    frame=lines,
    rulecolor=\color{black},
    xleftmargin=10pt,
    escapeinside={(*@}{@*)},
    showstringspaces=false,
    columns=flexible
}

\usetikzlibrary{fit}
\usepackage{nicematrix}
\usepackage{caption}

\microtypecontext{spacing=nonfrench}

\usepackage{enumitem}
\setitemize{noitemsep,topsep=0pt,parsep=0pt,partopsep=0pt}
\setenumerate{noitemsep,topsep=0pt,parsep=0pt,partopsep=0pt}
\setdescription{itemsep=0pt,topsep=0pt,parsep=0pt,partopsep=0pt,itemindent=0em,leftmargin=0.5em}

\newcommand{\captionfonts}{\bf \small}
\usepackage[normalem]{ulem}

\makeatletter  
\long\def\@makecaption#1#2{%
	\vskip\abovecaptionskip
	\sbox\@tempboxa{{\captionfonts #1: #2}}%
	\ifdim \wd\@tempboxa >\hsize
	{\captionfonts #1: #2\par}
	\else
	\hbox to\hsize{\hfil\box\@tempboxa\hfil}%
	\fi
	\vskip\belowcaptionskip}
\makeatother   

\newcommand{\squishlist}{
  \begin{list}{$\bullet$}{
    \setlength{\itemsep}{0pt}       \setlength{\parsep}{3pt}
    \setlength{\topsep}{3pt}        \setlength{\partopsep}{0pt}
    \setlength{\leftmargin}{1em}    \setlength{\labelwidth}{1em}
    \setlength{\labelsep}{0.5em} } }

\newcommand{\squishend}{
  \end{list} }

\newcommand{\squishenum}{
  \begin{enumerate}{}{
    \setlength{\itemsep}{0pt}       \setlength{\parsep}{3pt}
    \setlength{\topsep}{3pt}        \setlength{\partopsep}{0pt}
    \setlength{\leftmargin}{1em}    \setlength{\labelwidth}{1em}
    \setlength{\labelsep}{0.5em} } }

\newcommand{\squishenumend}{
\end{enumerate} }

\setlist[enumerate]{leftmargin=1em,topsep=1pt,itemsep=0pt}

\usepackage[textsize=tiny,textwidth=0.6in]{todonotes}
\newcommand{\exclude}[1]{}
\newcommand{\showComments}{yes}
\newcommand{\note}[2]{
    \ifthenelse{\equal{\showComments}{yes}}{{\vspace{-9em}\todo[color={#1!50}]{#2}\vspace{9em}}}{}
}

\newcommand{\ckpt}{shadow}
\newcommand{\Ckpt}{Shadow}

\newcommand{\sys}{{Checkmate}\xspace}
\newcommand{\pytorch}{PyTorch\xspace}
\newcommand{\ms}[1]{\SI{#1}{\milli\second}}

\makeatletter
\newcommand\footnoteref[1]{\protected@xdef\@thefnmark{\ref{#1}}\@footnotemark}
\makeatother

\usepackage{titlesec}
\titlespacing*{\section}{0pt}{7pt}{7pt}
\titlespacing*{\subsection}{0pt}{7pt}{7pt}
\titlespacing*{\subsubsection}{0pt}{7pt}{7pt}
\titlespacing*{\paragraph}{0pt}{1ex}{0.5ex}

\title{\textsc{\sys}: Zero-Overhead Model Checkpointing via \\ Network Gradient Replication}

\author{
{\rm
Ankit Bhardwaj\hyperref[equalcontrib]{\textsuperscript{*}} \qquad
Weiyang Wang\hyperref[equalcontrib]{\textsuperscript{*}} \qquad
Jeremy Carin \qquad
Adam Belay \qquad
Manya Ghobadi} \\\\
{Massachusetts Institute of Technology}}
\date{}

\begin{document}
\maketitle
\begingroup
\renewcommand\thefootnote{}\footnotetext[\value{footnote}]{\label{equalcontrib}\textsuperscript{*}Equal contribution.}
\endgroup

\begin{abstract}
    This paper presents Checkmate, a system that enables per-iteration checkpointing in DNN training without any training slowdown. The traditional approach to checkpointing requires a pause in training to copy model states to a separate location, allowing the state to be restored in the event of failure. This approach fundamentally has a tradeoff between the frequency of checkpoints and the cost of a failure. We avoid this tradeoff; our key insight is that in data-parallel training, all information necessary to create a checkpoint already exists in the network as gradients. Our core contribution is a new multicast abstraction that simultaneously delivers gradients to a separate CPU-based shadow cluster. The shadow maintains a checkpoint by applying those gradients to a copy of the model. Our evaluation shows that Checkmate performs per-iteration checkpointing with training throughput comparable to an ideal no-checkpoint baseline. \sys achieves $5$ to $34.5\times$ more frequent checkpointing compared to state-of-the-art checkpointing systems, resulting in $80\%$ to $97.1\%$ reduction in repeated work per failure. At the same checkpointing frequency, \sys delivers $1.3\times$ to $6.5\times$ throughput compared to other systems.
\end{abstract}

\section{{Introduction}} \label{sec:intro}

Today's deep neural networks (DNNs) have hundreds of billions of parameters, and training them requires tens of thousands of GPUs for months at a time~\cite{openai,rail-only}. At such a massive scale, failures are the common case. Meta's training of LLaMA3 encountered 419 failures~\cite{llama3}, and Alibaba reported failure rates as high as 43\% for its large training jobs~\cite{alibaba}.

The loss in progress caused by these failures is costly to operators. To help mitigate this cost, existing training frameworks periodically generate \textit{checkpoints}, a process where they \textit{interrupt} training to save model state to persistent storage~\cite{checkfreq,check-n-run,gemini,torch_dcp}. In the event of failure, training resumes from the most recent checkpoint, so that less work has to be recomputed.

Unfortunately, existing checkpointing systems have high overheads, so they leave operators with a dilemma: frequent state saving reduces lost work but slows down training through its regular interruptions, whereas infrequent state saving reduces interruptions, but more work has to be recomputed in the event of a failure.

\begin{figure}[t]
    \centering
    \includegraphics[width=0.75\columnwidth]{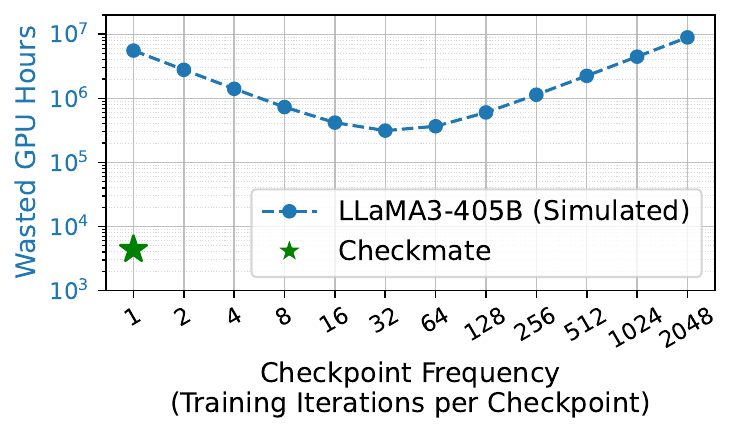}
    \caption{Total wasted GPU hours as checkpoint frequency changes, for LLaMA3-405B training. \normalfont Existing checkpoint frameworks trade off normal-case efficiency with repeated work in failure cases, while \sys breaks this tradeoff.}
    \label{fig:intro_recompute}
\end{figure}

Figure~\ref{fig:intro_recompute} illustrates this tradeoff using public data on the training of LLaMA3-405B. The total wasted GPU hours are plotted on the $y$-axis, defined as the sum of recomputation after failures and time spent on checkpointing, across different checkpoint frequencies~($x$-axis).
On the right side of the plot, infrequent checkpointing wastes GPU hours due to recomputation on failure. For example, Meta's prior work~\cite{check-n-run} used a 30-minute checkpoint interval (between 256 and 512 iterations in Figure~\ref{fig:intro_recompute}), resulting in 1.7 million wasted GPU hours, equating \$15 million at current cloud prices~\cite{google_cloud_pricing}. On the other hand, checkpointing frequently wastes GPU hours due to the time it stalls training. For instance, the left-most point of Figure~\ref{fig:intro_recompute} represents checkpointing every iteration. It minimizes repeated work but incurs high overall waste, as each checkpoint slows the normal iteration time by 28\%. Even at the best checkpoint frequency (every 32 iterations), the system still wastes over 300,000 GPU hours, costing \$3.3 million.\footnote{We estimate the {iteration time}, checkpointing overhead, and cost based on Meta's LLaMA3 technical report and public cloud GPU pricing with detailed derivations in Appendix~\ref{app:llama_iter} and~\ref{app:cost}.}  Existing checkpointing techniques~\cite{checkfreq,fastpersist,deepfreeze,gemini} reduce overheads but cannot break this fundamental tradeoff (detailed analysis in \S\ref{sec:eval:gpu_save}).

In this paper, we introduce \sys, a network-assisted checkpointing system that resolves this dilemma, enabling checkpointing every iteration without slowing down training. This is achieved through two key observations. First, model updates are deterministic: given a model state at time $t$ and its corresponding gradients, applying the optimizer step produces the state at time $t+1$, whether for training \textit{or} checkpointing. Second, in data-parallel training, these gradients are already computed and exchanged over the network every iteration. \sys captures these in-flight gradients and applies them to a prior checkpoint replica on a separate CPU cluster, reconstructing the latest model state without GPU involvement. Together, these insights enable per-iteration checkpointing with virtually no training overhead.

To realize this design, \sys introduces a \textit{multicast-update} framework that offloads checkpointing to the network and a dedicated cluster of CPU nodes. It replicates reduced gradients directly at programmable switches and \textit{multicasts} them to shadow nodes, which apply these \textit{updates} to independently maintained model replicas. The checkpointing process remains transparent to GPUs, allowing training to proceed without disruption. This allows \sys to substitute wasted GPU hours on recomputation with cheaper CPU hours on checkpointing.
For the aforementioned LLaMA3-405B model, the shadow cluster consumes only 166,000 CPU-node hours to enable per-iteration checkpointing, reducing wasted GPU time from over 300,000 hours to just 4,367 as shown in Figure~\ref{fig:intro_recompute}, cutting GPU waste by over 98\% and saving an estimated \$2.6 million in training costs.

\sys must meet three key requirements to function correctly and at scale: \textit{selecting} the gradients to replicate from the network, \textit{reliability} in delivering them, and \textit{timeliness} in applying updates fast enough to keep up with training.

For \textit{selection}, 
\sys faces the challenge that the network carries many types of traffic during each iteration. \sys must select the final version of the gradients exactly once per iteration.
We develop a lightweight tagging mechanism that marks final gradients precisely once per iteration for replication~(\S{\ref{sec:subsec:heartbeat}}). To ensure all tagged traffic reaches the shadow cluster, our system strategically places switches and shadow nodes in the network topology, guaranteeing all tagged gradients reach their multicast destination~(\S{\ref{sec:subsec:topology}}). This requires additional switch ports for connecting shadow nodes, a necessary overhead to ensure scalability.

For \textit{reliability}, \sys must ensure all tagged gradients reach the shadow cluster without packet loss or corruption, since losing gradient updates would result in invalid checkpoints.
It avoids congestion by using dedicated switch ports for shadow nodes, thereby isolating replication traffic from the training traffic.  
\sys enables Priority Flow Control (PFC), a widely adopted mechanism in training networks to handle transient receiver-side pressure.~(\S{\ref{subsec:network}})

For \textit{timeliness}, \sys must ensure that shadow nodes, despite being CPU-based, can keep pace with the GPU training rate.
\sys exploits the fact that each parameter's update is independent for popular optimizers and deterministically partitions the model state across the shadow cluster~(\S{\ref{subsec:storage_cluster}}). Scaling out the shadow cluster also ensures that flow control mechanisms are rarely triggered.

We build a small-scale prototype of \sys and evaluate it using popular vision and language DNN models on a 16-node testbed, comprising 12 training nodes equipped with NVIDIA A100 80 GB GPUs and 4 CPU-based {\ckpt} nodes, connected by 100 Gbps network interface cards (NICs).
\sys achieves per-iteration checkpointing with virtually no impact on training throughput, matching the no-checkpoint performance across all models. Compared to production systems like PyTorch DCP~\cite{torch_dcp} and state-of-the-art research prototypes like CheckFreq~\cite{checkfreq} and Gemini~\cite{gemini}, \sys achieves $5$ to $34.5\times$ more frequent checkpointing, resulting in $80\%$ to $97.1\%$ reduction in repeated work per failure. At the same checkpointing frequency, \sys delivers $1.3\times$ to $6.5\times$ throughput (\S\ref{sec:eval:comparison}). Our simulation shows that at 16K GPU scale, \sys saves 70,000 GPU hours over a 54-day run, even with a failure rate at $0.5\%$ of Meta's reported value~(\S{\ref{sec:eval:gpu_save}}). We will make our code public upon publication.

\section{Background}
\label{sec:bg}

\subsection{Overview of Distributed DNN Training}
\label{sec:bg:distributed}

Training large models requires splitting computations across multiple GPUs to manage large datasets. In data parallel (DP) training, each GPU holds a full copy of the model and processes different batches of training data in parallel. When models become too large to fit on a single GPU, pipeline parallelism (PP) is employed to divide the model into stages on different GPUs~\cite{megatronv2}. In practice, large-scale training often combines DP and PP to scale both models and datasets, where each pipeline stage is independently replicated across a group of GPUs as a DP group.

Each training iteration involves three phases: forward, backward, and optimizer step. In the forward pass, each DP group processes a batch of data. The backward pass computes gradients and synchronizes them with other DP groups. Finally, the optimizer uses these gradients to update model parameters. 
{Frameworks like PyTorch divide gradients into fixed-size buckets to facilitate more efficient communication. }

\paragraph{Gradient synchronization algo.}
GPUs perform gradient synchronization using an operation called \textit{AllReduce}. AllReduce typically consists of two stages: \textit{ReduceScatter} followed by \textit{AllGather}.
During ReduceScatter, each node splits its gradients into chunks, exchanges these chunks with other nodes, and reduces them (e.g., by summing or averaging). After this phase, each node holds a different chunk of the fully reduced result. In the following AllGather phase, each node sends its reduced chunk to another node. This process is repeated multiple times until every node has all the reduced chunks.

\subsection{Training Failures}
\label{sec:bg:checkpointing}

As model sizes increase and training scales grow, failures during the training process become more common. On each failure, the entire training job must restart. To avoid losing all training progress, practitioners periodically save the model state to a reliable medium, a process known as \textit{checkpointing}.

Existing checkpointing approaches use \textit{copy-persist} mechanism: they first copy model states (i.e., model parameters and optimizer states) out of GPU memory and then persist them to a reliable medium.
The most straightforward approach is to pause training and perform checkpointing \emph{synchronously}, where each GPU waits for the copy-persist operation to complete before resuming training, resulting in longer GPU stalls.

Recent work, such as CheckFreq~\cite{checkfreq}, Gemini~\cite{gemini}, and PyTorch Distributed Checkpointing~(DCP)~\cite{torch_dcp, torchsnapshot}, attempts to mitigate these stalls by making \textit{copy-persist} \textit{asynchronous}.
These systems overlap the copy step with forward and backward passes, as the model state is quiescent during this time. The persist step is then completed asynchronously before the start of the next checkpoint operation, reducing the overall stall time.
PyTorch Distributed makes further improvements by sharding checkpoints across training nodes, while CheckFreq dynamically tunes the checkpointing frequency to balance overheads and lost work on recovery.
To overcome limited disk bandwidth and high I/O latencies, Gemini~\cite{gemini} saves checkpoints to remote memory within the training cluster, using spare training network bandwidth to copy model states.

Although state-of-the-art approaches reduce training stalls, this paper shows that these schemes have fundamental overheads.
Specifically, we identify two primary overheads. First, the copy operation requires GPUs to clone states in parallel with training, which interferes with the training computations.
Second, the persist operation must finish before the start of the next checkpoint to prevent unbounded memory usage, which ultimately limits the checkpointing frequency.

\section{Motivation}
\subsection{Need for Per-Iteration Checkpointing}
\label{sec:subsec:need}

While checkpointing prevents a complete restart, progress loss is still inevitable, as any training progress made after the last checkpoint is lost and must be repeated on recovery. As models and cluster sizes grow, this repeated work becomes increasingly costly. For example, during LLaMA3-405B training on 16K GPUs, a single failure requires repeating 4,096 GPU-hours of computation on average when checkpointing every 30 minutes. Throughout the entire training, Meta reported 419 failures and wasted compute adds up to as much as \$15 million at current Google Cloud prices~\cite{google_cloud_pricing}.

With \textit{per-iteration} checkpointing, failure recovery incurs the absolute minimum repeated work. 
Checkpointing every iteration saves millions of dollars for a single job in large-scale training by reducing repeated work. However, existing systems introduce substantial checkpointing overhead, slowing down training and consuming additional resources during the checkpoint process itself. The following subsection quantifies the overheads in today's state-of-the-art systems.

\begin{figure}[t]
    \centering
    \includegraphics[width=\columnwidth]{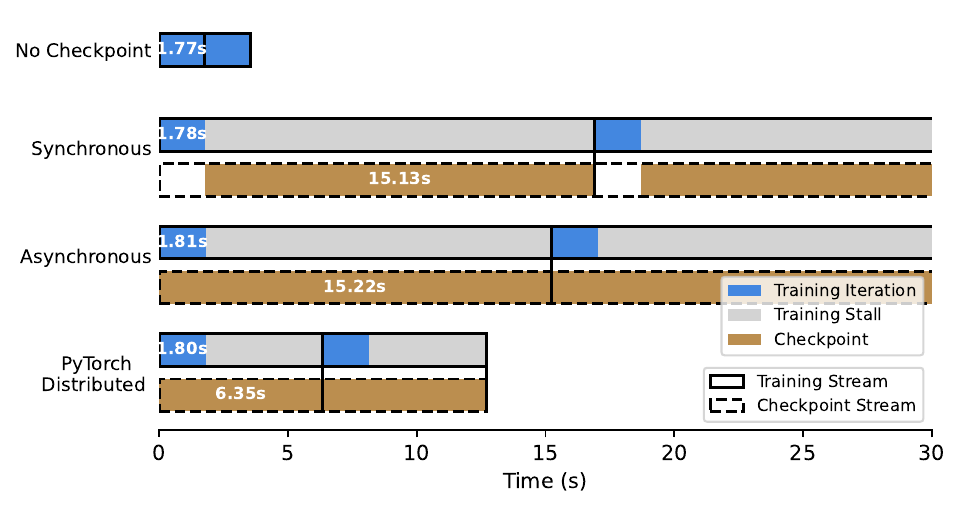}
    \caption{GPT3-XL training iteration times and stalls with various checkpointing approaches. \normalfont Existing approaches do not fully remove stalls when checkpointing per-iteration.}
    \label{fig:checkpoint_timeline}
\end{figure}

\subsection{Limitations of Existing Solutions}

Existing systems reduce checkpointing overhead but cannot eliminate it, especially at per-iteration frequency. The fundamental limitation lies in the \textit{copy-persist} model: copying from GPU to CPU memory introduces stalls due to the bandwidth gap between GPU and CPU, while persisting data to storage media adds further delays. With fast training iterations, these combined latencies prevent checkpointing from fully overlapping with computation, leading to unavoidable GPU stalls.

To quantify the overheads of existing systems, we compare four checkpointing techniques to illustrate this limitation. We measure the iteration time and checkpoint stalls for GPT3-XL with a batch size of 16 on a 4-node setup~(details in \S\ref{sec:eval:setup}).
Figure \ref{fig:checkpoint_timeline} highlights these measurements for two iterations, for comparison. The average iteration time for this setup is 1.77 seconds~(as shown in the top-most row in Figure~\ref{fig:checkpoint_timeline}).
The simplest strategy, synchronous checkpointing, performs the worst, causing a 9.5$\times$ slowdown by stalling training during the entire checkpoint.
Asynchronous checkpointing hides some of this time by overlaying the start of the checkpoint with training, but ultimately must still copy and persist the same total volume of data, causing a slowdown of 8.45$\times$.
PyTorch Distributed reduces the data volume each node needs to persist by sharding the checkpoint. Even still, we observe a 3.5$\times$ slowdown when sharded across four nodes. While smaller shards on a large cluster would reduce these slowdowns, our experimental results show that even with smaller shards, the overheads are still significant. Gemini aims to achieve per-iteration checkpointing, but our evaluations show that the training throughput slows by up to $3.47\times$ when checkpointing every iteration (\S\ref{sec:eval:comparison}).

When checkpointing per-iteration, current systems slow training across most model sizes. While further optimization of copy-persist is possible, we propose rethinking the problem by \textit{removing the GPU from the checkpointing process}.

\subsection{Pauseless Checkpointing Per-Iteration}
\label{sec:bg:opportunity}
To enable checkpointing every iteration without disrupting training, we seek to remove the GPU from involvement in checkpointing entirely. At first glance, this goal appears impossible, as the model state lives in GPU memory. However, this implication is only valid if one focuses solely on extracting checkpoint data directly from the GPU. 
Instead, we ask a different question: what data is already available during training that we can reuse for checkpointing rather than copying the entire model each time?

Our first insight is to \textit{reuse existing checkpoints}. Model updates are deterministic functions of the previous model states and the \textit{gradients} during training. Hence, we can apply the same optimizer step if we have the gradients and a copy of the prior model \textit{as a checkpoint}. This insight opens up \textit{incremental} update for per-iteration checkpointing. 

Our second insight is that these gradients are \textit{already in the network}. In DP training, GPUs exchange gradients every iteration to synchronize updates. Hence, rather than extracting data from the GPU, we tap into this existing data flow and replicate the gradients to a dedicated checkpoint cluster.

\sys combines these two ideas and introduces a \textit{multicast-update} framework for checkpointing. In \sys, we replicate the gradients \textit{in-network} using hardware multicast and stream them to a pool of CPU ``shadow” nodes. These nodes apply the same updates as the GPUs to maintain consistent model copies, enabling per-iteration checkpointing with no involvement from the training GPUs.

\section{\sys System Design}
\label{sec:design}

\begin{figure}[t]
    \centering
    \includegraphics[width=\columnwidth]{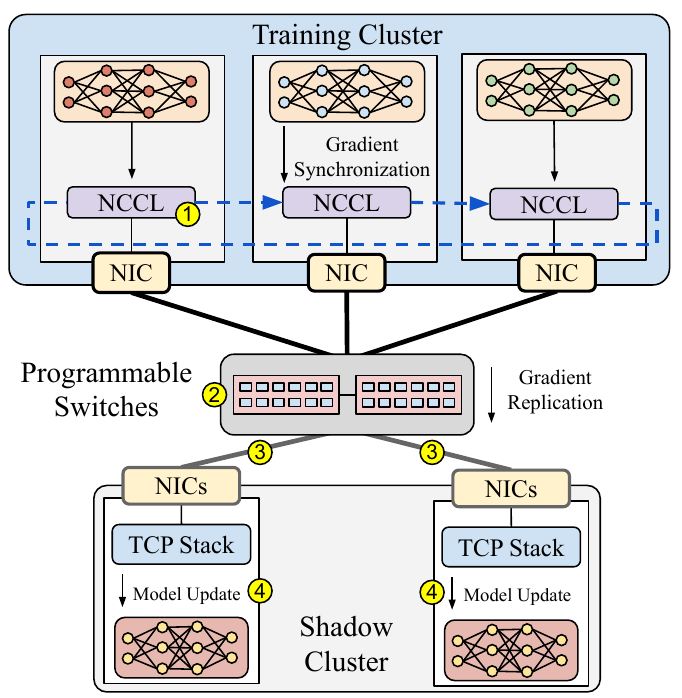}
    \caption{Overview of \sys's architecture. \normalfont The diagram highlights the interaction between training nodes, the network switch, and {\ckpt} nodes to enable pauseless checkpointing.}
    \label{fig:arch}
\end{figure}

\paragraph{Overview:} \sys enables pauseless per-iteration checkpointing. It \textit{selects} the gradients to replicate from the network, delivers them \textit{reliably} to the shadow nodes, and applies updates fast enough to keep up with training in a \textit{timely} manner.
It achieves this through a three-way collaboration across the training nodes, the network switches, and dedicated shadow CPU nodes. Figure~\ref{fig:arch} highlights \sys's main steps:

\begin{description}
    \item[{\small{\textcircled{1}}} Tagging gradients on training nodes.] \sys tags gradients before transmitting over the network, enabling the network data plane to distinguish them from other traffic.
    \item[{\small{\textcircled{2}}} Multicasting in the switch.] The switch multicasts tagged gradients to their training destination and the appropriate {\ckpt} nodes, while forwarding other packets normally. Priority Flow Control (PFC) ensures lossless delivery. 
    \item[{\small{\textcircled{3}}} Receiving gradient on {\ckpt} nodes.]  {\Ckpt} nodes receive the gradients and map them to the correct model layers.
    \item[{\small{\textcircled{4}}} Running optimizer step on {\ckpt} nodes.] Once all gradients for an iteration are received, {\ckpt} nodes run the optimizer step, updating their model and optimizer state.
\end{description}

Using these steps, \sys ensures that {\ckpt} nodes remain up to date with the training state, enabling failure recovery at a per-iteration granularity.

\subsection{Gradient Tagging on Training Nodes}
\label{sec:subsec:heartbeat}

To ensure correct and reliable operation, \sys must deliver each reduced gradient to the {\ckpt} cluster exactly once per iteration. Any duplication or loss would result in an inconsistent checkpoint, ultimately leading to a failed recovery.
This requirement is at odds with the behavior of standard collective algorithms, such as Ring AllReduce, where training nodes exchange the same gradient chunks multiple times over several communication rounds.
During the AllGather phase in Ring AllReduce, each node sends and receives a portion of the reduced gradients in each round, repeatedly transmitting the same data over the network.
These algorithms improve bandwidth utilization but complicate our attempt to extract a single, complete copy of the gradients from the network.

To accurately identify reduced gradients in the network, \sys tags them on the training nodes before transmitting.
While our current design is based on NCCL's most widely used Ring AllReduce algorithm, this tagging approach is generalizable to other AllReduce algorithms as well~(\S\ref{sec:discussion}).

The AllGather phase in Ring AllReduce exchanges final gradients $(n-1)$ times over the network in a single round for $n$ training nodes. A naive approach is to tag them in the first round. However, this leads to an $n$-way incast problem on the {\ckpt} node, where all $n$ training nodes simultaneously transmit their gradients to one {\ckpt} node, resulting in network congestion and data loss on {\ckpt} nodes.
Instead, \sys uses a heartbeat algorithm, where training nodes tag their gradient chunks evenly across $(n-1)$ AllGather rounds, distributing transmissions more uniformly.

\begin{figure}[t]
  \centering
  \begin{subfigure}[t]{1\linewidth}
    \centering
    \begin{tabular}{|
        >{\columncolor[HTML]{FFFFFF}}c |
        >{\columncolor[HTML]{FFCCC9}}c |
        >{\columncolor[HTML]{FFFFFF}}c |
        >{\columncolor[HTML]{FFFFFF}}c |}
      \hline
      Chunk-0 & Chunk-1 & Chunk-2 & Chunk-3 \\ \hline
    \end{tabular}
    \caption{Snapshot of the bucket for Rank-0 during round 0.}
    \label{tab:chunks}
  \end{subfigure}
  \vspace{5pt}
  \hfill
  \begin{subfigure}[t]{1\linewidth}
    \centering
    \footnotesize
    \begin{tabular}{c|ccc}
        \textbf{GPU} & \textbf{Round 0} & \textbf{Round 1} & \textbf{Round 2} \\
        \midrule
        0            & \textbf{C1}          & C0                   & C3                   \\
        1            & C2                   & C1                   & C0                   \\
        2            & C3                   & C2                   & C1                   \\
        3            & \textbf{C0}          & \textbf{C3}          & \textbf{C2}          \\
        \bottomrule
    \end{tabular}
    \caption{C1 and C0 are tagged in the first round, and C3 and C2 in the remaining two rounds. Each chunk is tagged exactly once.}
    \label{tab:allgather}
    \end{subfigure}%
    \caption{AllGather tagging example for a 4-GPU cluster.}
\end{figure}

\subsubsection{Heartbeat-based Tagging}

The heartbeat-based tagging mechanism is applied during $(n-1)$ AllGather rounds. In a ring of $n$ ranks, each rank sends one chunk of reduced gradients to its next neighbor and receives another chunk from its previous neighbor for $n-1$ rounds.
For example, Rank-0 starts iteration 0 holding Chunk 1 (see figure~\ref{tab:chunks}); it forwards that chunk to Rank-1 and simultaneously receives Chunk 0 from Rank $n-1$. After $n-1$ rounds every rank has gathered all $n$ chunks.
\sys minimally modifies the algorithm to tag chunks.
To ensure that each chunk is tagged exactly once, the algorithm only tags them on the boundary ranks: Rank-0 tags only its first chunk, and the last rank tags all its chunks before sending. This approach ensures full coverage with minimal overhead.

Figure~\ref{tab:allgather} illustrates how the algorithm distributes tagging across AllGather rounds on a 4-node cluster while also ensuring that all gradients are tagged precisely once.
Rank-0 tags only in the first round and Rank-3 in all the rounds, tagging C1, C0, C3, and C2, respectively.
This example shows a simplified version of the actual implementation. In practice, \sys handles multiple channels and connections, which are dynamically tuned by NCCL at runtime.
Importantly, \sys does not alter any other aspects of the algorithm or runtime, apart from introducing a lightweight 1-bit tagging step that has no measurable impact on performance.

While this approach reduces the likelihood of network incast on the {\ckpt} nodes, it does not eliminate it. In AllGather, $n$ gradient chunks are transmitted in only $(n-1)$ rounds. Consequently, one round inevitably involves two ranks transmitting data to a {\ckpt} node simultaneously~(C0 and C1 in the first round in Figure~\ref{tab:allgather} example).
To handle the bandwidth requirement, \sys equips each {\ckpt} node with two NICs, doubling their network bandwidth compared to training nodes. This setup allows {\ckpt} nodes to manage parallel transmissions from two training nodes without loss or congestion, ensuring \textit{reliable} reception of all gradients.
The dual-NIC approach does not significantly increase costs, as \sys requires only a few {\ckpt} nodes to keep up with training. We later show that only one {\ckpt} node is enough to handle large-scale models (e.g., GPT3-6.7B)~(\S\ref{sec:eval:storage}).

\subsubsection{Additional Metadata for Reassembling Buckets}
\label{sec:subsec:reassembly}

The heartbeat tagging algorithm ensures that the network receives a full copy of the tagged gradient. However, from the network flow's perspective, the algorithm tags non-consecutive data chunks in a continuous stream. This, coupled with parallel transmissions over multiple channels, makes gradient reassembly on the {\ckpt} nodes challenging.
To address this, the network layer maintains a separate sequence number for each channel, incrementing it only for tagged chunks. Then, it inserts the sequence number as a custom TCP option into the packets before sending them to the network.
The network switch mirrors each tagged packet and substitutes its original TCP sequence number with the channel-specific counter. This allows the {\ckpt} node to view the mirrored traffic on each channel as a single, continuous TCP stream.

\noindent Next, we focus on how shadow nodes maintain a checkpoint.

\subsection{{\Ckpt} Cluster}
\label{subsec:storage_cluster}

In this section, we describe the operations of the shadow cluster.
Each node in the cluster performs three tasks to maintain the checkpoint:
(1) it receives mirrored packets from the network and reassembles them into buckets,
(2) it then maps each bucket to its corresponding layer in the model, and
(3) it runs the optimizer step to maintain the checkpoint.

\subsubsection{Reassembling Buckets on {\Ckpt} Nodes}

\sys exploits the deterministic ordering guarantees of collective communication libraries (e.g., NCCL) to infer exactly which data chunks and offsets arrive on each channel.
This same determinism, which enables GPUs to run AllReduce and other collectives without branching, allows for predicting the sequence of incoming data for each channel on the {\ckpt} nodes.
\sys then uses this information to set up channels similar to those on the receive side of the training nodes, calculating the expected chunk sizes and offsets for each channel to sequence the incoming data and reassemble it into the original bucket.

Each {\ckpt} node allocates two sets of channels: one set to handle chunks from the first rank and another for the last, as tags are added only on these boundary ranks.
It then binds these channels to the two NICs in a round-robin manner to take full advantage of both NICs.
During the first AllGather round, when both first and last ranks are tagging data, both NICs operate at full capacity.
In subsequent rounds, each NIC operates at half capacity, avoiding scenarios where one NIC is idle while the other is near saturation.

\subsubsection{Mapping Bucket to Model Parameters}

After reassembling the buckets, each node must map these buckets to their corresponding layers in the model.
While this process can be complex, \sys handles the mappings in a way similar to the framework used on the training nodes. Training frameworks like PyTorch group parameters by bin-packing them, starting from the last model layer and working backwards to the first.
A model layer is mapped to a bucket until the bucket size is less than the maximum given size, such as 25MB in \pytorch DDP~\cite{ddp}. If a layer size exceeds the bucket size, it is mapped to a single dedicated bucket.
While \sys can maintain a different mapping to optimize the gradient capture from the network. The current design choice ensures that \sys easily integrates with existing training pipelines without additional overhead.
Besides the ease of integration, the approach also avoids the need for additional memory allocation. During the optimizer step, each model layer points to a specific offset in a bucket for the gradients without making additional copies.

\begin{lstlisting}[style=pythoncode, caption={Sample code for distributed training in PyTorch}, label={lst:ddp_example_training}, float=t]
model = DistributedDataParallel(dnn)
optimizer = Optimizer(model.parameters())
for batch, expected in zip(inputs, expected_outputs):
    optimizer.zero_grad() (*@\label{gpu:zero_grad}@*)
    output = model(batch) (*@\label{gpu:forward}@*)
    loss = loss_fn(output, expected) (*@\label{gpu:loss}@*)
    # Gradient computation and synchronization
    loss.backward() (*@\label{gpu:backward}@*)
    optimizer.step() (*@\label{gpu:step}@*)
\end{lstlisting}

\subsubsection{Updating the Checkpoint}

Once all the buckets are mapped, each shadow node runs the optimizer step to update the checkpoint (the parameter and optimizer states).
Each node runs the same loop as the training nodes, except {\ckpt} nodes skip the forward and backward passes and are replaced with the gradient receive logic from the switch.

Listing~\ref{lst:ddp_example_training} shows the code running on the training nodes. The code runs a forward pass~(line~\ref{gpu:forward}), calculates the loss~(line~\ref{gpu:loss}), and runs a backward pass for each batch~(line~\ref{gpu:backward}). Internally, the backward pass overlaps gradient computation and synchronization across different layers in the model. Lastly, the loop runs the optimizer step to update the model parameters and optimizer state~(line~\ref{gpu:step}).

Listing~\ref{lst:ddp_example_storage} shows the code running on the {\ckpt} nodes. It skips forward and backward passes altogether. Instead, {\ckpt} nodes wait for the gradients to arrive from the switch~(line~\ref{cpu:recv}) and run the optimizer step on CPUs to update the model states~(line~\ref{cpu:step}).

Although the optimizer update is computationally cheaper compared to forward and backward passes, a single {\ckpt} node can still become a bottleneck for large models. To address this, \sys allows scaling the optimizer step across multiple {\ckpt} nodes.

\begin{lstlisting}[style=pythoncode, caption={Sample code for {\ckpt} nodes}, label={lst:ddp_example_storage}, float=t]
model = DistributedDataParallel(dnn)
optimizer = Optimizer(model.parameters())
buckets = model.get_buckets()
while True:
    optimizer.zero_grad()
    buckets.recv()       (*@\label{cpu:recv}@*)
    optimizer.step()      (*@\label{cpu:step}@*)
\end{lstlisting}

\subsubsection{Scaling out the Optimizer Step}
\label{sec:scaling_out_the_optimizer}

\sys allows scaling the optimizer step across multiple nodes, ensuring {\ckpt} nodes keep pace with training nodes, even for large models.
We make no assumptions about the compute capabilities of {\ckpt} nodes, and allow users to configure the number of nodes based on model requirements. Before starting training, \sys profiles {\ckpt} nodes and configures the system for optimal performance.

For seamless scaling, \sys opts to support \textit{functional} optimizers, where the optimizer step for each parameter is deterministic and independent of the others.
Most optimizers, including SGD~\cite{sgd}, Adam~\cite{adam}, and AdamW~\cite{adamW}, are functional.
This property enables \sys to distribute the optimizer step across multiple nodes without affecting algorithmic correctness or introducing synchronization overhead.

Finally, distributing buckets across multiple nodes requires the switch to identify the correct {\ckpt} node for delivery. \sys encodes the {\ckpt} node ID in each packet for the switch to replicate gradients appropriately.

\paragraph{Consolidating Checkpoint State for Recovery:} Distributing the optimizer step adds an extra step in the recovery process. \sys uses a configurable timeout to consolidate shards into a complete checkpoint.
After consolidation, each {\ckpt} node serves as a checkpoint to the training nodes simultaneously, enabling a quick recovery process.

\subsection{Network Gradient Multicasting}
\label{subsec:network}

This section describes how the programmable switches, located between the training and checkpoint nodes, intercept and mirror packets tagged by the training nodes, and reliably deliver them to the checkpoint nodes.

\subsubsection{Switch Control-plane Setup}

The switch control plane must be configured before it can deliver gradients to the {\ckpt} nodes. It involves three steps:

First, the control plane is configured with the network addresses for the boundary ranks for each DP group. The switch uses this metadata to distinguish gradients from different shards.
For example, in a two-shard pipeline, both shards tag their gradients at the boundary ranks, and the switch uses those tags, along with the rank addresses, to tell them apart.
For each DP shard, the control plane creates \textit{protocol-independent multicast groups} for packet replication. For the first and last rank in a shard, the control plane identifies corresponding {\ckpt} nodes.
It creates a \textit{multicast group} with the next rank~(training) in the DP shard and the {\ckpt} nodes. These multicast groups are inserted into a match-action table, later used for multicasting tagged gradient packets.

Second, the switch requires a mapping between the shadow node ID, used for optimizer scale out, and {\ckpt} node IP addresses. The training nodes set the shadow node IDs to scale up the optimizer step, and the switch uses these mappings to update the destination on mirrored packets.

Lastly, the switch implements a minimal TCP server to emulate a parameter server for {\ckpt} nodes. So, before starting the training, each {\ckpt} node establishes TCP connections with the switch, which are later used to forward gradients over TCP streams. It drops ACK packets from {\ckpt} nodes.

\subsubsection{Intercepting and Multicasting Gradients} 
The data plane handles gradient multicasting statelessly. The switch performs regular L2 forwarding for untagged packets and match-action logic to process tagged ones. By carrying all necessary metadata within the packets, the switch avoids maintaining state, reduces complexity, and ensures adaptability to less complex switch architectures.

The ingress pipeline assigns a multicast group to the tagged packets based on control plane configurations and performs standard L2 forwarding for untagged ones.
At the end of the ingress pipeline, packets destined for {\ckpt} nodes undergo additional processing. The switch replicates the packets according to their multicast group. All packets are then handed off to the corresponding egress pipeline for transmission.

The egress pipeline rewrites the sequence number to the sequence number expected by the {\ckpt} nodes, stored in the TCP option field by the training nodes~(\S\ref{sec:subsec:reassembly}). It also updates the source and destination IPs for the {\ckpt} node’s TCP stream. Finally, the packet is forwarded to its destination.

\subsubsection{Lossless Packet Delivery} 
\label{subsec:lossless}

\sys requires all tagged gradients to reach their destination for a correct and consistent checkpoint. This requirement necessitates a lossless network for packet delivery.
Because training traffic saturates links only up to line rate and multicast mirrors the same data rate, \sys avoids introducing additional congestion by design.

Packet loss can still occur at the shared packet buffers within the switch or due to transient slowdowns in packet consumption at the {\ckpt} nodes. To avoid this issue, the system utilizes Priority Flow Control (PFC) for connections with {\ckpt} nodes.
In networks like RoCE, PFC is typically enabled system-wide, so no extra tuning is needed. In conventional datacenter networks, efficiently using PFC can be challenging \cite{pfc_storm}, but recent studies show that PFC works well for predictable patterns in large-scale training~\cite{llama3}.

\sys leverages protocol-independent multicast groups to facilitate efficient lossless delivery.
This feature enables configuring multicast with arbitrary fields in the packet header. Furthermore, modern switches enable all downstream devices in a multicast group to apply backpressure to the upstream source during congestion, effectively managing network bottlenecks and preventing packet loss for tagged traffic.
We tune PFC parameters, such as pause thresholds and buffer sizes, to align with workload requirements in both the training and the shadow cluster. This tuning maintains a lossless network, even under high traffic loads, ensuring reliable delivery of gradients for
checkpoint correctness.

\subsection{{Network Resource Planning}} \label{sec:subsec:topology}
This section analyzes \sys's network resource requirements and explains the topology choices to ensure scalability and performance isolation.

\paragraph{Network resource requirement.}
To replicate gradients in each iteration, \sys employs two multicast streams per DP group, regardless of the group's size. As a result, the total resource requirement scales linearly with the number of DP groups. For instance, the primary phase of Meta's LLaMA3-405B training utilized 128 DP groups. In this configuration, replicating gradients needs 256 multicast streams. Hence, \sys requires 256 additional switch ports, line-rate NICs, and corresponding transceivers to support replication. For a 16K-GPU setup, this amounts to less than $0.8\%$ more network resource of the training cluster, a modest overhead to ensure the reliability of replication traffic, and isolating them from the training path. Furthermore, multicast is a hardware-supported primitive on modern datacenter switches, enabling gradient replication at line rate. 

\paragraph{Where to put the shadow nodes.}
\sys must observe gradient synchronization traffic for all DP groups to facilitate replication. In hybrid-parallel training, DP synchronization traffic typically traverses the network even when faster interconnects like NVLink are present~\cite{llama3, rail-only}. Crucially, all inter-node traffic must pass through the Top-of-Rack (ToR) switches by definition, making ToR the ideal point for capturing and replicating DP traffic.
\sys connects shadow nodes to the ToR layer, enabling multicast-capable switches to observe and replicate gradient traffic directly. 

\paragraph{How to add additional ports.} There are several ways to accommodate this requirement: one option is to deploy leaf switches with a higher port density to absorb the additional load. Alternatively, some existing uplink ports can be repurposed to connect shadow nodes, slightly increasing the oversubscription ratio, a tradeoff made acceptable by recent work showing that large LLMs can be trained effectively even on spine-free topologies~\cite{rail-only}.

\section{Implementation}
\label{sec:impl}

We implement a prototype of \sys, comprising $\approx$~10000 lines of Rust, C, C++, Python, and P4 code for various components on the training nodes, {\ckpt} nodes, and the programmable switch. We describe the implementation of each component below.

\paragraph{Training nodes:} Figure~\ref{fig:stack_train} shows the software-stack on training nodes. Each node runs an unmodified \pytorch application while \pytorch itself has 50 lines of code changes, mainly to expose the bucketing interface to the user.
\sys uses NCCL v2.22.3 for collective communication, with only ten lines of code changes to tag the reduced gradients before sending them to the network stack.

The majority of our implementation efforts were focused on developing a NCCL network plugin~($\approx$~5000 lines of Rust code). NCCL allows using a custom network stack at runtime based on \texttt{NCCL\_NET\_PLUGIN} environment variable. If the variable is set, NCCL loads the plugin and uses it for communication.
This plugin is used to send and receive messages between the nodes~\cite{net-plugin}. The plugin handles NCCL API calls and adds tags and sequence numbers to the reduced gradients before sending them to the underlying TCP library.

The plugin uses libtpa, a DPDK-based TCP stack~\cite{libtpa}, for network transport. We modify libtpa to support custom sequence numbers and packet tagging~(with a DSCP bit).
Our plugin achieves up to 98.4 Gbps bus bandwidth in NCCL AllReduce on a 100~Gbps NIC, matching the performance of NCCL's InfiniBand implementation. The plugin's modular design also allows using it with other NCCL-based libraries.

\begin{figure}[t]
    \centering
    \footnotesize
    \begin{subfigure}[t]{0.49\columnwidth}
    \centering
    \begin{tikzpicture}
        \foreach \y/\color in {0/blue!20, 0.4/green!20, 0.8/yellow!20, 1.2/red!20, 1.6/orange!30} {
            \draw[fill=\color] (0,\y) rectangle (3.8,\y+0.4);
        }
        \node at (1.8,1.8) {\pytorch Training Application};
        \node at (1.8,1.4) {\pytorch $\Delta$};
        \node at (1.8,1.0) {NCCL $\Delta$};
        \node at (1.8,0.6) {Pluggable Net-Plugin $\star$};
        \node at (1.8,0.2) {DPDK TCP Stack (libtpa) $\Delta$};
    \end{tikzpicture}
    \caption{\scriptsize Training.}
    \label{fig:stack_train}
    \end{subfigure}
    \begin{subfigure}[t]{0.49\columnwidth}
    \centering
    \begin{tikzpicture}
        \foreach \y/\color in {0/blue!20, 0.4/green!50, 0.8/red!20, 1.2/gray!20, 1.6/orange!50} {
            \draw[fill=\color] (0,\y) rectangle (3.8,\y+0.4);
        }
        \node at (1.8,1.8) {\pytorch Checkpoint Agent $\star$};
        \node at (1.8,1.4) {\pytorch Sharded Optimizer $\star$};
        \node at (1.8,1.0) {\pytorch $\Delta$};
        \node at (1.8,0.6) {Rust Bucketing Interface $\star$};
        \node at (1.8,0.2) {DPDK TCP Stack (libtpa) $\Delta$};
    \end{tikzpicture}
    \caption{\scriptsize {\Ckpt} cluster.}
    \label{fig:stack_storage}
    \end{subfigure}
    \caption{Software-stack running on training and {\Ckpt} nodes. \normalfont $\star$ indicates components we implement, while $\Delta$ indicates components we apply minor changes only.}
    \label{fig:stack}
\end{figure}

\paragraph{{\Ckpt} nodes:} Figure~\ref{fig:stack_storage} shows the software-stack on {\Ckpt} nodes. We implement the checkpointing agent using the same high-level \pytorch code as the training nodes, but without forward and backward passes. Instead, it receives gradients from our custom bucketing interface to retrieve and store them.
The bucketing interface internally uses the DPDK-based TCP stack to communicate with the switch. The network stack is fast and saturates two 100 Gbps NICs.

An interesting implementation challenge was efficiently copying gradients from intermediate network buffers to \pytorch, particularly for bucket sizes ranging from 25~MB to 500~MB in LLMs. To overcome this, we develop a custom memory copying mechanism that leverages AVX-512 streaming instructions and parallelizes the operation across multiple idle CPU cores, achieving an 8~$\times$ speedup over the default Rust \texttt{memcpy}. This optimization significantly improves performance until bandwidth becomes the limiting factor.

We implement these functionalities using $\approx$~2000 lines of Rust code to leverage Rust's safety and performance features while avoiding Python's Global Interpreter Lock (GIL) issues. We wrap the Rust code with a Python interface using PyO3~\cite{pyo3} to interact with \pytorch. We implement the partitioned optimizer with $\approx$~1000 lines of Python code to scale out the optimizer on the {\Ckpt} nodes.

\paragraph{Switch:} The switch is responsible for routing packets between the {\Ckpt} and training nodes. We implement the data plane functionality with $\approx$~800 lines of P4~\cite{p4} and run it on a 32-port Tofino-1 Switch~\cite{tofino}. In addition, we implement the control plane using 1000 lines of Python code to register the replication stream, create multicast groups, and setup the TCP server. The control plane configures the switch using P4Runtime's gRPC API~\cite{grpc}.

\section{Evaluation}
\label{sec:eval}

This section evaluates the performance, correctness, and scalability of \sys across various models and system scales. We begin by describing our experimental setup (\S\ref{sec:eval:setup}), followed by a comparison against state-of-the-art checkpointing systems in terms of throughput and checkpoint frequency (\S\ref{sec:eval:comparison}). We then assess the shadow cluster's scalability, including optimizer step timing and the scaling behavior of CPU-based checkpoint nodes (\S\ref{sec:eval:storage}, \S\ref{sec:eval:sscale}). Next, we validate the correctness of checkpointing (\S\ref{sec:eval:weights}) and show that gradient multicasting introduces no measurable network overhead (\S\ref{sec:eval:network}). Finally, we quantify \sys's GPU-hour savings at scale under varying cluster sizes and failure rates (\S\ref{sec:eval:gpu_save}).

\begin{table}[t]
    \centering
    \footnotesize
    \NiceMatrixOptions{xdots/line-style = {densely dotted, blue}}
    \begin{NiceTabular}{@{}lcc:c@{}}
        \toprule
        \textbf{Model}           & \textbf{Parameters} & \textbf{Parallelism}   & \textbf{Min. CPU-nodes}    \\
        \midrule
        ResNet50~\cite{resnet}   & 25.6M               & 12 DP                  & 1  \\
        ResNet152~\cite{resnet}  & 60.2M               & 12 DP                  & 1  \\
        ViT-H-14~\cite{vit}      & 633.5M              & 12 DP                  & 1  \\
        GPT2-1.5B~\cite{gpt2}    & 1.5B                & 12 DP                  & 1  \\
        GPT3-XL~\cite{gpt3}      & 1.3B                & 12 DP                  & 1  \\
        GPT3-6.7B~\cite{gpt3}    & 6.7B                & 12 DP                  & 1  \\
        LLaMA2-7B~\cite{llama2}  & 7B                  & 2 PP $\times$ 6 DP     & 2  \\
        LLaMA2-13B~\cite{llama2} & 13B                 & 2 PP $\times$ 6 DP     & 2  \\
        LLaMA3-8B~\cite{llama3}  & 8B                  & 2 PP $\times$ 6 DP     & 2  \\
        \bottomrule
    \end{NiceTabular}
    \caption{Models and configurations used in the evaluation.}
    \label{tab:models}
\end{table}

\subsection{Experimental Setup}
\label{sec:eval:setup}

Our evaluation uses a total of 16 machines, including 12 training nodes equipped with Nvidia A100 80~GB GPUs, and up to 4 Intel Xeon Gold 5420+ {\ckpt} nodes with 28 cores each. All of these machines are connected through a 32-port Tofino-1 programmable switch operating at 100~Gbps.
The training machines are set up with CUDA 12.6, \pytorch 2.6.0, and modified NCCL v2.22.3 with tagging functionality, and the {\ckpt} nodes run the same \pytorch version.

\begin{figure*}[t]
    \centering
    \begin{subfigure}[t]{\textwidth}
        \centering
        \includegraphics[width=0.9\columnwidth]{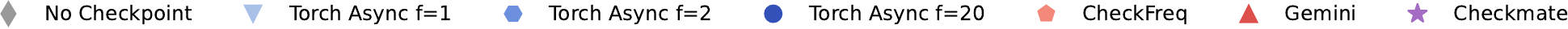}
    \end{subfigure}
    \begin{subfigure}[t]{\textwidth}
        \centering
        \includegraphics[width=\textwidth]{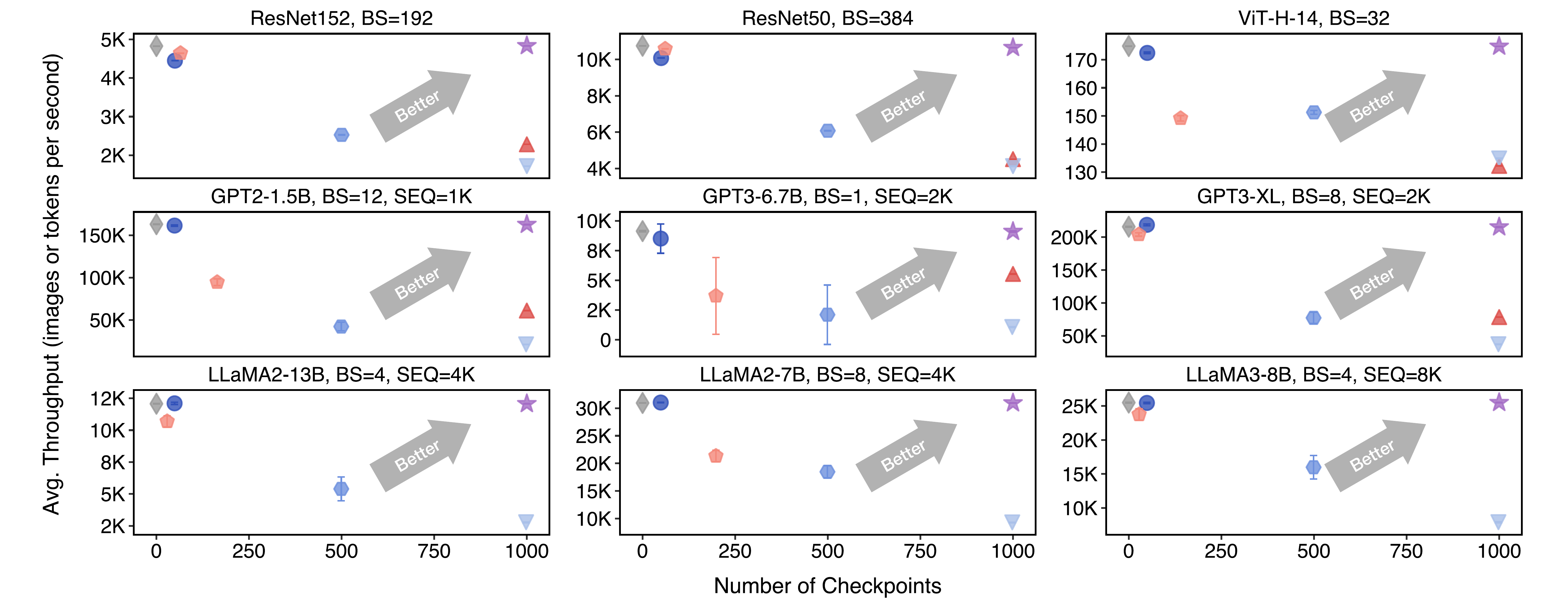}
    \end{subfigure}
    \caption{Training throughput and number of checkpoints for different systems. \normalfont \sys checkpoints 5 to 34.5$\times$ more compared to CheckFreq and achieves $1.32\times$ to $2.7\times$ throughput than per-iteration checkpointing in Gemini on average.}
    \label{fig:tput_nsnap_combined}
\end{figure*}

We evaluate \sys using various models, including ResNet and ViT for vision tasks, and GPT and LLaMA for language tasks. Smaller vision and GPT models are trained using pure DP, while each LLaMA DP replica is split 2-way via pipeline parallelism. Table~\ref{tab:models} shows the details for each model.
We train vision models on the ImageNet dataset~\cite{imagenet} with FP32 precision, while language models are trained on the C4 dataset~\cite{c4} with BF16 precision.
All the models are trained using the AdamW optimizer with a cosine annealing learning rate schedule.
The vision models use an unmodified \texttt{torchvision} library for the training. For language models, we extend TorchTitan~\cite{torchtitan} to support hybrid data and pipeline parallelism, allowing evaluation across a wider set of models.

\begin{figure*}[t]
    \centering
    \begin{subfigure}[t]{\textwidth}
        \centering
        \includegraphics[width=0.8\columnwidth]{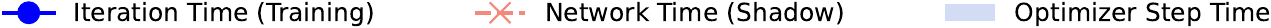}
    \end{subfigure}
    \begin{subfigure}[t]{\textwidth}
        \centering
        \includegraphics[width=\textwidth]{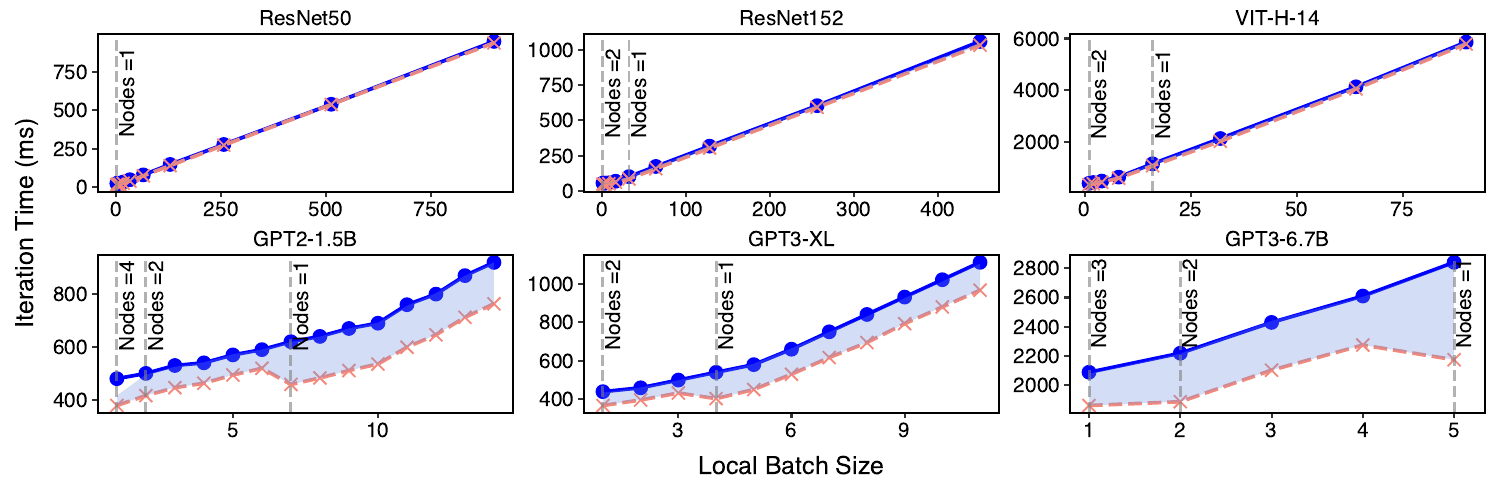}
    \end{subfigure}
    \caption{Impact of {\ckpt} node scaling on checkpointing time for different models. \normalfont With a single {\ckpt} node, \sys achieves checkpointing times as low as \ms{25} for smaller models~(ResNet50), and completes in under \ms{600} even for LLMs~(GPT2-1.5B and GPT3-XL).}
    \label{fig:storage_scaling}
\end{figure*}

\subsection{Comparison with Other Systems}
\label{sec:eval:comparison}

We evaluate \sys against state-of-the-art production systems and research prototypes for models listed in Table~\ref{tab:models}. To isolate checkpointing effects, \sys and the No Checkpoint baseline utilize our DPDK-based network stack, whereas other systems use NCCL's default RoCE-v2 stack. We set up in-memory \texttt{nullfs}~\cite{nullvfs} across all systems to prevent persistent storage writing from becoming a bottleneck. We run 1000 training iterations for each configuration.

Figure~\ref{fig:tput_nsnap_combined} shows the result. The $x$-axis in each sub-figure represents the number of checkpoints over 1000 iterations, and the $y$-axis indicates throughput. The title on each sub-figure displays the model, local batch size, and sequence length, if applicable. 

Each marker represents the throughput and checkpoint count for a given system. We also report the saved repeated work upon failure, derived from the checkpointing frequency. For example, if a given system checkpoint occurs every $6$ iterations, then upon failure, it needs to repeat $3$ iterations of work on average. Since \sys checkpoints per iteration, it always incurs $0.5$ iterations of repeated work for each failure. Therefore, \sys needs to do $(3-0.5)/3 = 83\%$ less repeated work compared to this base system.

\paragraph{No Checkpoint Baseline.}
The No Checkpoint baseline serves as an upper bound on throughput, indicating the maximum achievable training speed when checkpointing is disabled. \sys matches this baseline across all workloads, achieving virtually identical throughput. While achieving this performance, \sys also produces checkpoints every iteration, resulting in minimal repeated work in the event of a failure. This establishes the zero-overhead nature of \sys.

\paragraph{Torch Async~\cite{torch_dcp,torchsnapshot}.}
The next comparison system is the asynchronous checkpointing in PyTorch Distributed Checkpoint~(Torch Async).
Besides asynchronously copying the checkpoint state to local CPU memory and writing it to a local \texttt{nullfs}, Torch Async shards checkpoint states across training nodes, overall reducing the overhead.
We vary the checkpointing frequency ($f$) to evaluate trade-offs between checkpoint frequency and training throughput overhead for this system.

At $f=20$, Torch Async yields a low overhead, only 5\% slower in throughput compared to \sys on average. However, Torch Async needs to perform $95\%$ more repeated work for each failure due to producing checkpoints at only 1/20 frequency of \sys. On the other hand, \sys outperforms Torch Async at higher checkpointing frequencies. For instance, \sys achieves $2.2\times$ and $6.5\times$ throughput on average for vision and language models, respectively, when checkpointing per iteration. 
\sys outperforms PyTorch DCP on both fronts: zero checkpoint overhead and minimized repeated work.

\paragraph{CheckFreq~\cite{checkfreq}.}
On the high level, CheckFreq shares the same goal as \sys: maximizing checkpoint frequency while minimizing overhead. However, CheckFreq still relies on the copy-persist framework and performs such optimization by profiling the checkpoint overhead and tuning the checkpoint frequency accordingly. It performs asynchronous checkpointing and optimizes the copy step by copying intermediate states to GPU VRAM when the models are small. 

\sys consistently surpasses CheckFreq in both throughput and checkpoint frequency, regardless of model size. For instance, for ResNet152 (Figure~\ref{fig:tput_nsnap_combined} top left), where CheckFreq copies to GPU memory before persisting, \sys achieves $1.05\times$ throughput and $15.3\times$ more frequent checkpointing, indicating $91.5\%$ less repeated work for each failure. For LLaMA2-13B (Figure~\ref{fig:tput_nsnap_combined} bottom left), a model with high GPU memory utilization, \sys achieves $1.1\times$ throughput while checkpointing $34.5\times$ more frequently ($97.1\%$ less repeated work per failure). 

This result also illustrates the difficulty in choosing the right checkpoint frequency with profiling in practice: for GPT3-6.7B, LLaMA2-7B, and ViT-H-14, CheckFreq's initial profiling underestimated the overhead of checkpointing, resulting in more checkpoints but more overhead compared to the other models. \sys avoids this problem by offloading checkpointing to the network and the shadow cluster.

\paragraph{Gemini~\cite{gemini}.}
Like \sys, Gemini seeks to achieve per-iteration checkpointing by checkpointing into the remote CPU memory of the training cluster using GPUDirect-RDMA, avoiding storage stalls. It shards checkpoint states across training nodes, replicates these shards for reliability, and schedules checkpoint traffic to interleave with training without interfering with computation. In this evaluation, Gemini used the default replication factor of one. Note that Gemini employs DeepSpeed ZeRO-3, which adds an extra AllGather per iteration and increases the base iteration time compared to the other systems in our evaluation. Specifically, we exclude Gemini's results for LLaMA models, as the difference in parallelization (DP+PP hybrid vs. ZeRO-3) makes it difficult to draw fair comparisons.

\sys achieves higher throughput than Gemini across all models. On average, \sys delivers $2.2\times$ throughput while maintaining the same checkpoint frequency for the vision and GPT models. Since Gemini relies on the training network to transmit data, models like ResNet with fast iteration provide less opportunity to overlap checkpointing traffic and gradient transmission, hindering its performance, especially for small batch sizes. Its overhead will also increase with the replication factor, presenting a tradeoff in performance and reliability. \sys avoids these problems by utilizing a dedicated network and CPU resources.

\paragraph{Summary.}
Across all models, \sys consistently delivers per-iteration checkpointing at no cost to training throughput. As a result, \sys achieves the best normal-case performance and the least repeated work upon failure.

\begin{figure*}[t]
    \centering
    \begin{minipage}{.32\textwidth}
        \includegraphics[width=1.0\columnwidth]{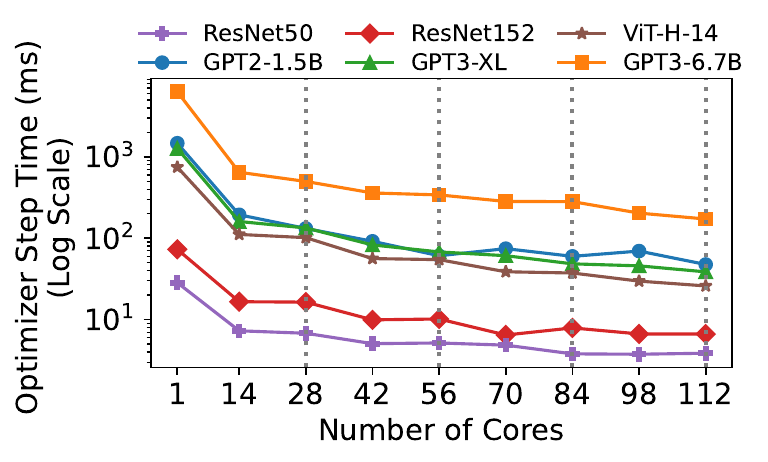}
        \caption{\normalfont \sys's optimizer step time scales linearly with the number of nodes.}
        \label{fig:sscale}
    \end{minipage}
    \hspace*{1mm}
    \begin{minipage}{.32\textwidth}
        \includegraphics[width=1.0\columnwidth]{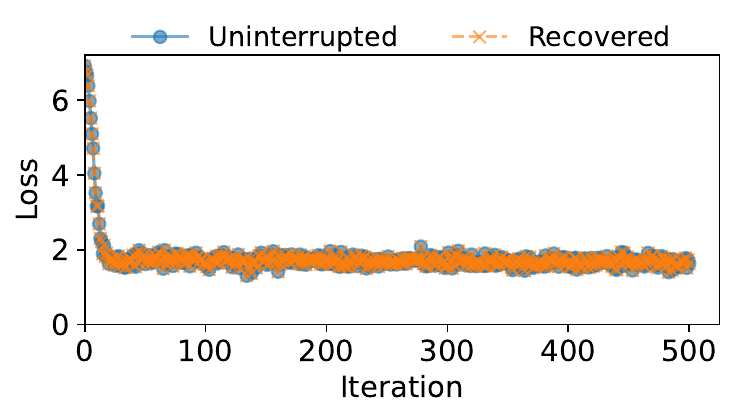}
        \caption{\normalfont The uninterrupted and recovered models converge identically (ResNet152).}
        \label{fig:equality}
    \end{minipage}
    \hspace*{1mm}
    \begin{minipage}{.32\textwidth}
        \includegraphics[width=1.0\columnwidth]{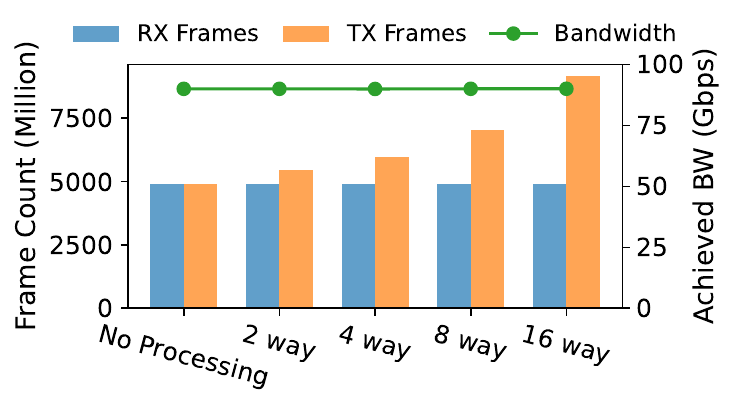}
        \caption{\normalfont Achieved bandwidth and frame count for different replication factors.}
        \label{fig:network_rep_speed}
    \end{minipage}
\end{figure*}

\subsection{How Fast can \sys Update?}
\label{sec:eval:storage}

The CPU-only {\ckpt} nodes must complete their optimizer step before the GPU-based training nodes begin the next one to keep pace with the GPU training nodes. We stress-test this timing constraint by sweeping the local batch size from one to the point of GPU memory saturation across both vision and GPT models.
Smaller batch sizes shorten the training iteration time, pressuring the {\ckpt} nodes to finish more quickly. In contrast, larger batch sizes increase the iteration time, but a smaller number of {\ckpt} nodes must be used to take full advantage of them.
To push the limits even further, this experiment uses AdamW, a widely used but computationally demanding optimizer.
For every model and batch size, we present the minimum number of nodes required to complete the optimizer step and the time taken for each configuration. Additionally, we measure the average iteration time and the time spent by {\ckpt} nodes on pulling gradients and applying the optimizer step for each configuration.

Figure~\ref{fig:storage_scaling} shows the results. The solid line depicts the total iteration time, and the dashed line indicates the time spent pulling gradients. The shaded region shows the time spent applying the optimizer step. The $x$-axis represents the local batch size, and the $y$-axis shows the iteration time in milliseconds. Vertical dotted lines mark the number of {\ckpt} nodes used for each configuration. We run each experiment for 200 iterations and average the results to minimize transient effects.

As expected, the iteration time increases with batch size for all models~(solid blue line). Interestingly, the time spent by {\ckpt} nodes pulling gradients (yellow dotted line) also increases with batch size, even though the total gradient size remains constant for a given model. This is because larger batch sizes lead to longer gradient computation times on the training nodes. During this time, {\ckpt} nodes remain idle, waiting for gradients to arrive, which increases the time spent pulling gradients from the network.

For the vision models~(top row in Figure~\ref{fig:storage_scaling}), the ratio of time spent by the {\ckpt} nodes pulling gradients to the optimizer step time is low, resulting in a minor shaded region. ResNet50 and ResNet152, being smaller models, have shorter optimizer step times. On the other hand, ViT-H-14 has a longer optimizer step time; the iteration time remains the dominant factor, which keeps the shaded region minimal.

For language models~(bottom row in Figure~\ref{fig:storage_scaling}), the optimizer step time is significantly longer than for vision models due to their higher compute requirements, resulting in a larger shaded region. Despite this, the {\ckpt} nodes keep up with the training nodes, often requiring only one server for larger batch sizes. For example, \sys uses a single server for GPT2-1.5B, GPT3-XL, and GPT3-6.7B, with batch sizes of 7, 4, and 5, respectively.
When a server is removed (e.g., at the batch size 7 for GPT2-1.5B), the shaded region expands, reflecting longer optimizer step times due to fewer available cores; however, {\ckpt} does not become a bottleneck, as the total {\ckpt} time consistently remains below the solid line.

\subsection{Optimizer Scaling across Multiple Nodes}
\label{sec:eval:sscale}

In this experiment, we analyze the impact of the number of CPU cores on the optimizer step time across various models. We vary the number of cores and measure the time it takes to run the optimizer step. Each {\ckpt} node has 28 cores, and we add an additional {\ckpt} node for every 28 cores to partition and distribute the optimizer step workload. We wrap each model with our stateless optimizer and run optimizer steps over randomized gradients.

Figure~\ref{fig:sscale} demonstrates that the optimizer time decreases sharply as we add more cores, even within a single {\ckpt} node. For instance, the optimizer step time for GPT3-6.7B decreases from \ms{6000} with a single core to \ms{500} with 28 cores.
The step time decreases almost linearly as we add more {\ckpt} nodes, with each vertical dotted line representing the addition of a new {\ckpt} node. For the same example, with 56 cores, the step time decreases to \ms{350}, and with all 112 cores, it decreases to \ms{180}.

For most models in our evaluation, two {\ckpt} nodes are sufficient to keep up with training. For example, with GPT3-6.7B, the iteration time for a local batch size of 4 is greater than \ms{2000}, but even with half the cores on a single {\ckpt} node, the optimizer step time is less than \ms{1000}. This highlights that even for the large model, \sys only needs a few {\ckpt} nodes, equipped with a typical datacenter server CPU, to keep up with the training nodes.

\subsection{Checkpoint Correctness}
\label{sec:eval:weights}

We investigate whether the model and optimizer states are mathematically equivalent across the training and {\ckpt} clusters. Ensuring this is crucial to verify that the training process remains unaffected by the checkpointing mechanism.
To verify this, we train a vision model~(ResNet152) over the ImageNet dataset for 500 iterations over four training nodes and one {\ckpt} node. We repeat the training process twice using the same random seed and identical hyperparameters. In the first run, training proceeds uninterrupted, and we record the training loss after each iteration. In the second run, we intentionally halt training during every second iteration, restoring the model weights and optimizer states from the {\ckpt} nodes before resuming the next iteration.

Figure~\ref{fig:equality} shows the loss for ResNet152. The solid line represents the uninterrupted run, and the dashed line represents the interrupted run. The lines overlap completely, indicating that the training loss over time is identical. This observation confirms that \sys's checkpointing mechanism preserves the mathematical equivalence of the training process.

To further validate these findings, we perform a third experiment. In this case, we compare the model weights, biases, and optimizer states, such as momentum and learning rate, up to the 8th decimal place between the training and {\ckpt} clusters after each iteration. The comparison shows no discrepancies, reinforcing the conclusion that the \sys maintains checkpointing correctness.

\subsection{Network Overheads of Multicasting}
\label{sec:eval:network}

\sys uses the network switch to efficiently multicast tagged gradients without slowing training. Tofino-1 has a dedicated Packet Replication Engine (PRE) capable of multicasting at full line rate. We want to ensure that the switch can operate at line rate, even with a higher replication factor.

To verify this, we run the NCCL AllReduce test on 1~GB of data for 200 iterations across four training nodes and replicate the tagged gradients to 2 to 16 additional switch ports. We measure the AllReduce bus bandwidth and the total transmitted frames, as counted by the switch. For comparison, we evaluate the baseline throughput without packet replication.

Figure~\ref{fig:network_rep_speed} shows the bus bandwidth and transmitted frames. The RX frame count and the total data transmitted from the GPUs remain the same across all replication factors.
Given that the switch only replicates tagged packets, an increase in TX does not proportionally increase the network load on the switch. Even at 16-way replication, the switch only transmits 1.9$\times$ the number of frames it receives.
Note that the switch does not replicate all the packets, but only the tagged ones.
In all cases, the AllReduce bus bandwidth remains constant.

\begin{figure}[t]
    \centering 
    \includegraphics[width=0.9\columnwidth]{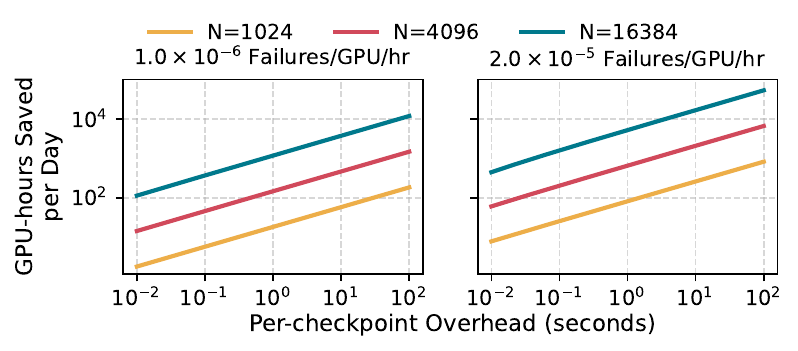}
    \caption{GPU hours saved per day by \sys compared to traditional checkpointing systems across varying cluster scales, checkpoint overheads, and GPU failure rates. \normalfont \sys yields increasing savings at larger scales, remains effective under low checkpoint overhead, and remains beneficial at low failure rates.}
    \label{fig:gpu_saved}
\end{figure}

\subsection{GPU Savings at Scale}
\label{sec:eval:gpu_save}

The prior section evaluated \sys's feasibility and performance on a local testbed. This section extends the analysis to larger scales and varying failure rates with simulation.

Figure~\ref{fig:gpu_saved} shows the GPU hours saved by \sys compared to traditional checkpointing systems. We assume a training workload similar to LLaMA3-405B, with an iteration time of 4.58 seconds (Appendix~\ref{app:llama_iter}). Based on the checkpoint overhead, we apply the optimal checkpoint frequency that minimizes the expected waste of GPU hours (Appendix~\ref{app:cost}).

For each subplot, the $x$-axis sweeps checkpoint overhead values to reflect a range of implementations, from synchronous to various asynchronous checkpointing, while the $y$-axis illustrates the expected GPU hours saved per day by \sys. Each line corresponds to a different cluster size, and each subplot represents different GPU failure rates. The right-hand side subplot uses Meta's reported failure rate from LLaMA3 training at $2.0\times 10^{-5}$ failures per GPU-hour.

Three key observations emerge from this analysis. First, \sys yields greater savings as the cluster size increases. \sys saves 16 times more GPU hours at 16,384 GPUs compared to 4,096 GPUs for both subfigures. This stems from the quadratic increase in wasted work with system scale, as derived in Appendix~\ref{app:cost}. Second, \sys provides meaningful benefits even under low checkpoint overhead. The right-hand side figure shows that assuming just 10~ms overhead per checkpoint (lowest point on $x$-axis), \sys still saves 448 GPU-hours per day at the 16,384-GPU scale (blue line).  Over a 54-day training period, this results in more than 24,000 GPU-hours saved. Third, \sys remains effective even if failure rates improve significantly. The left-hand side figure demonstrates that at a failure rate of $10^{-6}$ failures per GPU-hour (0.5\% of Meta’s observed rate), \sys still saves nearly 70,000 GPU-hours across a 54-day training run.

\section{Related Work}
\label{sec:related}

\sys builds on a rich body of work in distributed training, model checkpointing,
and in-network computation. Throughout the paper, we have discussed prior efforts on checkpointing; in this section, we discuss the most relevant work in the rest of each area.

\paragraph{Optimizer offloading and sharding.}

Systems like DeepSpeed ZeRO-Offload~\cite{deepspeed,zero-3}, leverage CPU offloading to reduce GPU memory usage and utilize CPU computing, enabling the training of larger models. In addition to offloading computations, these systems commonly use CPU memory to store optimizer states~\cite{gemini,checkfreq}, activations~\cite{nemo}, and weights. Systems like DeepSpeed~\cite{deepspeed} and NVIDIA NeMo~\cite{nemo} effectively use CPU memory for these purposes to optimize resource usage.
Techniques such as ZeRO, FSDP, and OSDP~\cite{zero-3,fsdp,osdp} shard the optimizer state across multiple devices mainly to reduce memory usage.
\sys builds on optimizer offloading and sharding techniques to enable efficient checkpointing.

\paragraph{Fault-tolerance using redundant computation.}

Redundant computation represents another class of fault-tolerance techniques in distributed training. Systems like Bamboo~\cite{bamboo}, Oobleck~\cite{oobleck}, and ReCycle~\cite{recycle} use redundant computation to handle stragglers and failures. While effective, these approaches incur high computational and communication overhead due to duplicate computations. In contrast, \sys minimizes overhead by using minimal CPU resources to maintain shadow models and optimizer states, which are only accessed during recovery, offering a lightweight and efficient alternative for fault tolerance.

\paragraph{Optimizing DNN training with in-network computation.}

In-network computing optimizes distributed DNN training by offloading computations to network infrastructure. Systems like SwitchML~\cite{switchml}, ATP~\cite{atp}, SHARP~\cite{sharp}, and PANAMA~\cite{panama} use programmable switches to perform gradient aggregation, reducing communication overhead.
While these systems focus on optimizing gradient communication, \sys complements them by capturing gradients during training and transmitting them to a shadow cluster for checkpointing. Unlike aggregation systems, \sys does not require specialized switch hardware, such as floating point units, making it more widely applicable.

\paragraph{Lossless delivery for datagrams.}

\sys relies on lossless packet delivery for correct gradient replication. While this requirement appears stringent, large-scale production systems have demonstrated the feasibility of deploying lossless networks in real-world settings. For instance, Meta's DNN training clusters~\cite{roce_metascale} and Microsoft's Azure Storage~\cite{azure_storage_rdma} achieve reliable lossless delivery at scale. In Meta's LLaMA3 training infrastructure, mechanisms like DCQCN are disabled, and PFC serves as the sole method for flow and congestion control, even at scales involving 16K GPUs~\cite{llama3}. Although we do not evaluate \sys at such a scale, it leverages the same PFC mechanism, indicating its potential for scalability. Additionally, recent research has shown that lossless delivery is feasible even in high-demand scenarios. For example, BFC~\cite{bfc} demonstrates reliable lossless transmission in data center networks with high-degree incast scenarios (e.g., 2000-to-1), further supporting the deployability of lossless delivery mechanisms for large-scale systems. 

\section{Discussion}
\label{sec:discussion}

\paragraph{Applicability to other AllReduce algorithms.} Section~\ref{sec:subsec:heartbeat} describes \sys's gradient replication algorithm based on Ring-AllReduce. However, \sys is not limited to this algorithm. AllReduce operations generally follow one of two patterns: a ReduceScatter followed by an AllGather (e.g., Ring-AllReduce, as used in our implementation) or a Reduce followed by a Broadcast (e.g., the double binary tree algorithm~\cite{nccl_double_binary_tree}). In both cases, the network transmits reduced gradients, enabling \sys to identify and replicate the required information to the {\ckpt} cluster.

\paragraph{\sys and FSDP training.}

Techniques like FSDP~\cite{fsdp} and ZeRO~\cite{zero-3} replace the AllReduce with a ReduceScatter for gradients and an AllGather for model parameters, which appears to be incompatible with \sys. However, \sys can extend its multicasting abstraction to support these strategies. Specifically, \sys multicasts model parameters in the AllGather step in FSDP to {\ckpt} servers for checkpointing, ensuring parameters are recoverable even if optimizer states are not directly captured.

For optimizer states, if the update step is linear to the gradient, we invert the update equation and solve for the gradient with new parameters, subsequently updating the optimizer state. For optimizers that track the second moment, such inversion yields a quadratic equation and, thus, two possible solutions for the gradient. In these cases, \sys can still function with minor modifications, such as transmitting additional metadata during the AllGather step to resolve ambiguities. We recognize that fully adapting \sys to support FSDP and ZeRO requires a comprehensive evaluation of the trade-offs, which we leave as future work.
\section{Conclusion}
\sys introduces a novel approach to checkpointing that leverages a multicast operation to efficiently capture gradients from the network and offload checkpointing to a scalable shadow cluster. This design eliminates training pauses or slowdowns. \sys achieves $5$ to $34.5\times$ more frequent checkpointing compared to state-of-the-art checkpointing frameworks, resulting in $80\%$ to $97.1\%$ reduction in repeated work per failure. At the same checkpointing frequency, \sys delivers $1.3\times$ to $6.5\times$ throughput compared to other systems.
\label{sec:conclusion} 
\section*{Acknowledgments}
We thank Om Chabra, Gohar Irfan Chaudhry, Josh Fried, Pouya Hamadanian, Ben Holmes, Zain Zhenyuan Ruan, and Anton A. Zabreyko, as well as other members of the Network and Mobile Systems (NMS) and Parallel and Distributed Systems (PDOS) research groups, for their helpful feedback.
This research was supported by NSF SHF-2107244, NSF CAREER-2144766, NSF PPoSS-2217099, NSF CNS-2211382, NSF FuSe-TG-2235466, Sloan fellowship FG-2022-18504; and by ACE, one of seven centers in JUMP 2.0, a Semiconductor Research Corporation (SRC) program sponsored by DARPA.

\bibliographystyle{plain}
\bibliography{paper.bib}

\newpage
\cleardoublepage
\appendix
\section{Iteration and Checkpoint Time of LLaMA3}
\label{app:llama_iter}
In the LLaMA3 technical report~\cite{llama3}, Meta did not specify the training iteration time for each phase, rendering it hard to estimate the repeated work. However, with other published data, it is possible to calculate their training iteration time, which we report below. 

We estimate the training iteration time by calculating the FLOPs (floating-point operations) for a single LLaMA training iteration. We start by estimating the FLOPs required for a single LLaMA-style Transformer model forward pass employing Grouped Query Attention (GQA). The key operations considered include attention projections, attention computations, feed-forward networks (FFN), rotary positional embeddings (RoPE), and the final language modeling head projection. We provide a succinct overview of each component and the formula for estimating its computational cost. Table~\ref{tab:notation} summarizes the notation we use in our derivation. 

\begin{table}[ht]
\small
\centering
\begin{tabular}{cl}
\hline
Symbol & Meaning \\
\hline
$b$ & Batch size \\
$s$ & Sequence length \\
$L$ & Number of Transformer layers \\
$h$ & Hidden dimension \\
$f$ & Feed-forward (FFN) dimension \\
$v$ & Vocabulary size \\
$a$ & Total number of attention heads for queries (Q) \\
$g$ & Number of groups for keys (K) and values (V) in GQA \\
\hline
\end{tabular}
\caption{Notation used in FLOPs calculation}
\label{tab:notation}
\end{table}

\paragraph{Attention Projections (QKV).} Each Transformer layer computes linear projections to obtain queries (Q), keys (K), and values (V). We calculate the cost as:
\begin{align*}
    \text{FLOPs}_{\text{QKV}} &= 2 \left( b \cdot s \cdot h^2 + 2 \cdot b \cdot s \cdot h \cdot (g \cdot a) \right)
\end{align*}

\paragraph{Attention Computation.} The dot-product attention and the subsequent multiplication with V incur:
\begin{align*}
    \text{FLOPs}_{\text{Attn}} &= 4 \cdot b \cdot s^2 \cdot h
\end{align*}

\paragraph{Attention Output Projection.} Projecting attention outputs back to the hidden dimension costs:
\begin{align*}
    \text{FLOPs}_{\text{AttnOut}} &= 2 \cdot b \cdot s \cdot h \cdot (g \cdot a)
\end{align*}

\paragraph{Feed-Forward Network (FFN).} Each FFN layer includes two linear transformations:
\begin{align*}
    \text{FLOPs}_{\text{FFN}} &= 4 \cdot b \cdot s \cdot h \cdot f
\end{align*}

\paragraph{Rotary Positional Embedding (RoPE).} Positional embeddings add a computational overhead, approximated by:
\begin{align*}
    \text{FLOPs}_{\text{RoPE}} &= 2 \cdot b \cdot s \cdot h
\end{align*}

The total FLOPs per Transformer layer combine all these components:
\begin{align*}
    \text{FLOPs}_{\text{Layer}} &= \text{FLOPs}_{\text{QKV}} + \text{FLOPs}_{\text{Attn}}\\
      &+ \text{FLOPs}_{\text{AttnOut}} + \text{FLOPs}_{\text{FFN}} + \text{FLOPs}_{\text{RoPE}}
\end{align*}

\paragraph{Accumulation over Layers.} Summing across $L$ layers:
\begin{align*}
    \text{FLOPs}_{\text{TotalLayers}} &= \text{FLOPs}_{\text{Layer}} \cdot L
\end{align*}

\paragraph{Final LM Head Projection and word embedding.} Projecting hidden states to the vocabulary dimension, and projecting the vocabulary into the hidden dimension costs:
\begin{align*}
    \text{FLOPs}_{\text{Vocab}} &= 4 \cdot b \cdot s \cdot h \cdot v
\end{align*}

\paragraph{Total FLOPs.} The final estimated FLOPs for a single forward pass are:
\begin{align*}
    \text{FLOPs}_{\text{Total}} &= \text{FLOPs}_{\text{TotalLayers}} + \text{FLOPs}_{\text{Vocab}}
\end{align*}

The total FLOPs of an iteration is roughly three times the FLOPs of one forward pass, since the backward pass performs the same amount of operations for both the activation gradients and the weight gradients. Note that the LLaMA3 technical report specified that activation checkpointing is disabled. Therefore, the backward pass does not enclose another forward pass of computation. 

We validate our formulation by computing the total FLOP for LLaMA training with Meta's reported numbers and comparing it to the $3.5\times10^{25}$ FLOPs training budget specified by LLaMA. Accounting for all three phases in pretraining, our formula yields a total training FLOP of $ 3.49\times 10^ {25}$, which closely matches the published number.

With this estimation, the training iteration time is 
\begin{equation}
    \text{Iteration time} = \frac{\text{FLOPs}_{\text{Total}}}{\text{Achieved FLOP per GPU}\times\text{Total GPUs}}
\end{equation}

For the major phase of LLaMA3-405B training with a batch size of 16M, sequence length of 8192, and total GPU count of 16384, Meta achieved 400~TFLOPs per-GPU~\cite{llama2}. Using our formulas above, the estimated iteration time is 4.58s.

To estimate the checkpointing time, we use numbers reported in the LLaMA3 technical report, where Meta mentions that their checkpoint storage cluster has a sustainable throughput of 2~TB/s. For the 405B parameter model, the total checkpoint size is 
\begin{align*}
    &\text{405B parameters} \times \\
    &(\text{2 bytes per parameter} + \text{4 bytes per optimizer state}) \\
    &= 2.43\text{~TB}
\end{align*}
Hence, the checkpoint process takes 1.2s.

\section{Cost Analysis}
\label{app:cost}
In this section, we analyze the operational cost of \sys. We seek to understand the operational scheme of \sys: given a model and a checkpoint overhead, how does scale and failure rate impact \sys's effectiveness in saving?

\subsection{Cost Modeling}

We utilize cloud pricing for GPUs and CPUs and model the associated costs. Table~\ref{tab:cost_anal} summarizes the variables we use in this derivation.

\begin{table}[ht]
\small
\centering
\begin{tabular}{cl}
\hline
Variable & Meaning \\
\hline
$\lambda$ & Failure rate of GPUs (per hour) \\
$N$ & Number of GPUs \\
$D$ & Total training duration without failure \\
$f$ & Checkpoint frequency\\
$t$ & Iteration time\\
$\omega$ & Checkpoint stall time\\
$C$ & Number of CPU nodes required for \sys\\
$P_G$ & GPU price (\$/GPU/hour)\\
$P_C$ & CPU node price (\$/CPU-Node/hour)\\
\hline
\end{tabular}
\caption{Notation used in cost calculation.}
\label{tab:cost_anal}
\end{table}

In existing checkpointing systems, the total wasted GPU hours consist of the repeated work from GPU failure and the checkpointing overhead. Checkpointing happens every $ft$ time. Assuming a constant failure rate, a failure occurs uniformly between checkpoints. Therefore, the expected repeated work is half of the checkpoint interval for each failure $\frac{ft}{2}$. For an $N$-GPU system, assuming GPU failures are IID, then the failure follows a binomial distribution. Therefore, the expected number of failures is $\lambda ND$. Thus, the expected wasted total GPU hour during the entire training duration is $\frac{1}{2}\lambda N^2Dft$.
 
The checkpoint overhead depends on the checkpointing frequency $f$. Assume each checkpoint stalls the training with time $\omega$. For simplicity, we assume the per-checkpoint overhead is a constant in this derivation, though we note that as checkpoint frequency goes up, this overhead increases due to the copy and persist step contending each other more. Then, the total checkpoint overhead throughout the entire training is $\frac{ND}{ft}\omega$. The sum of both terms provides the total wasted hour:
\begin{equation}
    \text{Wasted}_{SOTA}(f) = ND\left(\frac{1}{2}\lambda Nft + \frac{\omega}{ft}\right)
\end{equation}
This is the equation for the GPU-hour curve in Figure~\ref{fig:intro_recompute}. For a per-hour GPU price $P_G$, the total cost of wasted GPU hours, as a function of checkpoint frequency, is  
\begin{equation}
    \text{Cost}_{SOTA}(f) = P_GND\left(\frac{1}{2}\lambda Nft + \frac{\omega}{ft}\right)
\end{equation}

For \sys, we require $C$ CPU-nodes to update the checkpoint together with training consecutively. Therefore, \sys will spend $D C$ total CPU-node hours and achieve a per-iteration checkpoint. \sys still faces, on average, half an iteration of repeated work, which amounts to $\frac{1}{2}N^2D\lambda t$ GPU hours. Hence, \sys's total cost of keeping checkpointing plus the cost of wasted GPU hours is
\begin{equation}
    \text{Cost}_{\text{\sys}} = \frac{1}{2}P_G\lambda N^2D + P_CDC
\end{equation}
\sys becomes the optimal solution for checkpointing when 
\begin{equation}
    \text{Cost}_{\text{\sys}} < \text{Cost}_{SOTA}^*
\end{equation}
Where $\text{Cost}_{SOTA}^*$ is the minimum cost of a conventional system for \textit{any} checkpoint frequency. Taking the derivative of $\text{Cost}_{SOTA}(f)$, we find the optimal checkpoint frequency that minimizes cost. The frequency is also at least one, since it is infeasible to checkpoint fractional iterations. Therefore, the optimal checkpoint frequency is 
\[   
f^* = 
     \begin{cases}
       \sqrt{\frac{2\omega}{\lambda N t^2}}, &\qquad \sqrt{\frac{2\omega}{\lambda N t^2}}\geq 1\\
       1, &\qquad \text{otherwise.} \\ 
     \end{cases}
\]
This gives the minimum cost 
\[   
\text{Cost}_{SOTA}^* = 
     \begin{cases}
       P_GND\sqrt{2\omega\lambda N}, &\qquad \sqrt{\frac{2\omega}{\lambda N t^2}}\geq 1\\
       P_GND\left(\frac{1}{2}\lambda Nt + \frac{\omega}{t}\right), &\qquad \text{otherwise.} \\ 
     \end{cases}
\]

To estimate the total cost of repeated work, we utilize the public cloud pricing of H100 SXM5 GPUs, the same GPU used in Meta's LLaMA3 training cluster, on Google Cloud, leveraging Google's cloud platform cost estimator~\cite{google_cloud_pricing}. At the time of writing, H100 GPU costs \$11.06 per hour, while a CPU node with 32 cores and 128~GB of DRAM costs \$1.28 per hour.

We calculate the CPU time for \sys with LLaMA3 assuming 128 CPU servers, each handling two multicasting streams. Since LLaMA3 training run for 54 days, the total CPU-node hours are $54\text{ days}\times 128\times \text{ days} \times 24\text{ hours/day}=166\text{K CPU-node hours}$.

\end{document}